\documentstyle[psfig,epsf]{mn}

\title [Quasar Environment in the Context of Large Scale Structure at $z \sim 0.3$] {Quasar Environment in the Context of Large Scale Structure at \boldmath ${z \sim 0.3}$}

\author [I. K. S\"{o}chting et al.]
{
Ilona K. S\"{o}chting,$^{1}$
Roger G. Clowes,$^{1}$
Luis E. Campusano$^{2}$ \\
$^{1}$ Centre for Astrophysics, University of Central Lancashire,
Preston PR1 2HE (E-mail: imachura@uclan.ac.uk) \\
$^{2}$ Observatorio Astron\'omico Cerro Cal\'an, Departamento de
Astronom\'{\i}a, Universidad de Chile, Casilla 36-D, Santiago, Chile
}

\date{Accepted 2001 Xxxxxxxx xx. Received 2001 Xxxxxxxx xx; in original form
2001 Xxxxxxxx xx}

\def\gs{\mathrel{\raise1.16pt\hbox{$>$}\kern-7.0pt
\lower3.06pt\hbox{{$\scriptstyle \sim$}}}}
\def\ls{\mathrel{\raise1.16pt\hbox{$<$}\kern-7.0pt
\lower3.06pt\hbox{{$\scriptstyle \sim$}}}}

\begin {document}

\maketitle

\begin {abstract}
We are looking at quasar environment in the context of large-scale structure
(LSS) --- a new approach, giving a more informed interpretation of
quasar-galaxy correlations. This paper presents our first results for a sample of $z \sim 0.3$ quasars. We are using Voronoi
tessellation applied in colour ($B_J - R$) slices for the detection of galaxy
clusters and the minimal spanning tree (MST) to delineate the large-scale structure. This new
cluster-detection method allows us to find reliably galaxy clusters at $z <
0.3$ from SuperCOSMOS measurements of UK Schmidt plates. By reconstructing the large-scale structure in a relatively narrow redshift band ($0.2 < z < 0.3$), we show that quasars follow the large-scale structure traced by galaxy clusters. None of the quasars in our radio-quiet sample is located in the central area of a galaxy cluster. Two quasars, found in a very rich environment, are actually located between two very close galaxy clusters, consistent with results on $z \sim 1$ quasars suggesting that cluster mergers may be involved in one of the quasar formation mechanisms.
\end {abstract}

\begin {keywords}
galaxies: clusters: general -- quasars: general -- cosmology: large-scale structure of Universe.
\end {keywords}

\section {INTRODUCTION}

Quasar environments have been studied on different scales, from those of the
host galaxy to those of the quasar-cluster cross-correlation function. Yee \&
Green (1984) found that there
is a difference in richness between the environments of radio-quiet and radio-loud
quasars (RQQs and RLQs respectively). This result has been confirmed by
Ellingson, Yee \& Green (1991), showing that RQQs are much less frequently situated
in very rich galaxy clusters than RLQs. Confirmation came also from Hintzen, Romanishin \& Valdes (1991), who found a significant excess of galaxies in the vicinity of
RLQs, and from Boyle \& Couch (1993), who found no significant excess of faint
galaxies in association with RQQs. Hutchings, Crampton \& Johnson (1995) obtained a different result; they observed the
galaxy environment of RLQs and RQQs and concluded that there is no
significant difference in the richness. Since the RQQs in that sample were
members of a large quasar group (LQG) (Crampton, Cowley, \& Hartwick 1989,
1987), this result raised the question of whether the environments of
members of LQGs could be different. However, some of the more recent results for quasars in general
imply that the richness of the environments of radio-loud and radio-quiet
quasars could indeed be very similar (e.g. Saxton, Hall \& Turner 1999, Wold et
al. 2001), whereas others still support the scenario of RLQs inhabiting more
dense groups of galaxies than RQQs (Hutchings et al. 1999, Teplitz et
al. 1999, S\'{a}nchez \& Gomz\'{a}les-Serrano 1999). 

The above results have been obtained using galaxy counts or the quasar-galaxy cross-correlation  function within small radii ($< 1h^{-1}$ Mpc) around the quasars. The interpretation of the results is not a straightforward task, because of the erasure of directional information. For example, quasars residing centrally in poor clusters would produce similar statistics to those on the priepheries of very rich clusters.

Originally, the twofold environment of quasars was interpreted as supporting the connection
of RLQs with elliptical host galaxies and RQQs with spirals
(e.g. Ellingson et al. 1991). However, in the past few years evidence has
emerged that RQQs may have diverse host galaxies. In a sample studied by
Bahcall et al. (1997), for example, more RQQs have been found in elliptical
hosts than in spirals. The result has been confirmed by Boyce et al. (1998),
and by McLure et al. (1999), who found that, just like radio-loud quasars,
essentially all RQQs with luminosities greater than $M_{R}=-24$ reside in massive ellipticals. In
many cases the host galaxies show traces of continuing interactions or past
mergers (e.g. Bahcall et al. 1997, Boyce et al. 1998, Hutchings et
al. 1999). Even if the host galaxies appear in the optical passbands to be
isolated and apparently undisturbed, they are found to exhibit continuing or
remnant tidal HI disruptions, indicating galactic encounters or mergers (Lim \& Ho 1999).

The emerging picture is that RLQs have elliptical host galaxies and prefer a
cluster environment whereas the RQQs inhabit various host galaxies in a wide
range of environments. At present we have no information on how the close
environment ($< 1h^{-1}$ Mpc) of quasars is related to the large-scale structure traced by
galaxy clusters, and if indeed the local enhancements of the density in the
vicinity of quasars can be interpreted as galaxy clusters. This prompted us to introduce a new
approach and look at the environments of quasars in the context of large-scale structure (LSS).

\setcounter{table}{0}
\begin{table*}
\begin{minipage}{120mm}
\caption{Low redshift quasar sample. $X$ and $Y$ are the coordinates of the quasars in arcminutes from the the field centre. The notation AQD/UVX means AQD quasars re-discovered by CUQS.}
\label{tab1}
\begin{tabular}{ccrrccl}
 RA (J2000)  & DEC (J2000) &   X    &   Y    &   z   &  $B_J$ & Survey  \\
             &            & arcmin & arcmin &       &   mag  &         \\
\hline
 10 33 56.8  &  +05 48 30 &  $-$130  &   64   & 0.376 &  19.3  & AQD/UVX \\
 10 34 47.9  &  +05 16 16 &  $-$117  &   32   & 0.331 &  19.2  & AQD/UVX \\
 10 34 58.3  &  +05 52 32 &  $-$114  &   68   & 0.297 &  19.5  &   UVX   \\
 10 35 06.0  &  +06 01 41 &  $-$112  &   77   & 0.245 &  18.9  &   UVX   \\
 10 36 15.7  &  +03 19 15 &   $-$95  &  $-$85   & 0.389 &  18.6  &   AQD   \\
 10 36 26.3  &  +04 54 35 &   $-$92  &   10   & 0.382 &  18.3  &   AQD   \\
 10 38 28.2  &  +04 45 35 &   $-$62  &    1   & 0.286 &  18.9  &   UVX   \\
 10 38 31.1  &  +06 09 12 &   $-$61  &   85   & 0.290 &  18.1  &   UVX   \\
 10 38 57.9  &  +04 48 39 &   $-$54  &    4   & 0.357 &  19.0  &   UVX   \\
 10 44 06.4  &  +03 14 50 &    23  &  $-$89   & 0.267 &        &  Keable \\
 10 44 52.0  &  +03 52 51 &    34  &  $-$51   & 0.213 &  17.2  &  Keable \\
 10 45 04.8  &  +04 34 01 &    37  &  $-$10   & 0.395 &  19.5  &   UVX   \\
 10 45 35.1  &  +03 41 22 &    45  &  $-$63   & 0.355 &  19.3  &   UVX   \\
 10 46 10.6  &  +03 50 31 &    54  &  $-$54   & 0.363 &  19.2  &   UVX   \\
 10 50 42.8  &  +04 17 39 &   122  &  $-$27   & 0.230 &  19.4  & AQD/UVX \\
\hline
\end{tabular}
\end{minipage}
\end{table*}

At low redshift, the LSS can be conveniently recognised as density peaks in
multidimensional parameter space (ra, dec, magnitude, colour) (Gladders \& Yee 2000). Wide
coverage (over $6 \times 6$ deg$^{2}$) is required to ensure the isotropic
detection of structures on scales $\sim 60h^{-1}$~Mpc. In this paper
we describe our first results from this new approach. The
quasar and galaxy samples used in this study are described in Section 2. The
galaxy cluster detection method, including a short introduction to Voronoi
tessellation, is outlined in Section 3. In Section 4 we have compiled the
cluster catalogue from which the LSS has been reconstructed. The results of
the study of the environments of quasars are contained in Sections 5 and
6. The advantages of the new approach and the implications of our results are
discussed in Section 7. Throughout this paper we adopt a standard
cosmological model with $\Omega = 1$, and $H_0
= 100h$ km~s$^{-1}$~Mpc$^{-1}$.

\section {DATA}

Our quasar sample consists of 15 low-redshift ($z < 0.4$) quasars from two
surveys covering most of ESO/SERC field 927 (1950 field centre
10$^h$~40$^m$~00$^s$ $+$05$^\circ$~00$'$~00$''$). Five quasars are from the
AQD survey which uses Automated Quasar Detection (AQD --- Clowes 1986; see
also Clowes, Cooke \& Beard 1984). The spectroscopic observation of
high-grade candidates were carried out at the CTIO 4-m telescope (Clowes \&
Campusano 1991; Clowes \& Campusano 1994; Clowes, Campusano \& Graham
1999). Eight quasars come from the Chile-UK Quasar Survey (CUQS ---
Newman et al. 1998; Newman 1999) which uses the ultraviolet excess (UVX)
method for selection. The spectroscopic observations were carried out at the Las Campanas 2.5m Du Pont telescope. Two quasars have come from Keable (1987). The CTIO quasars have
faint limits ($B_J \sim 20.5$ mag) and sparse sampling whereas the LCO quasars have bright limits ($B_J \sim 19.5$ mag)
and complete sampling. The quasar sample will be incomplete,
but, nevertheless, free of any spatial bias. Table 1 contains the basic information about the quasars in our sample.

Objects for the determination of the density distribution have been drawn
from UKST $B_J$ and $R$ plates covering $\sim 6.3 \times 6.3$ deg$^2$ of
sky. The plates have been digitized on the SuperCOSMOS measuring machine at
the Royal Observatory Edinburgh. Only objects recognized on both plates
became a part of the final catalogues. This process eliminates virtually all
of the spurious detections in the individual plate scans (due to satellite
and meteor trails, emulsion noise, etc.). The blue plates (J10128 and J10063) have been calibrated by applying $3^{rd}$ degree polynomial fitting to deep (limiting $B_J \sim 21.5$) CCD sequences from Graham (1997). For the faint end ($B_J > 21.5$) a linear extrapolation has been applied. The $R$ plates (R18775 and R10071) have been calibrated in a similar way but using Keable (1987) data as secondary calibrators. The photometrically calibrated data set is limited at $B_{J}=22.5$ and $R=21.5$.

Star-galaxy separation is not reliable at such faint limits, but for the
purpose of this study faint stars can be treated as forming a uniform background. Jones
et al. (1991) found that the distribution of star counts in cells (small areas of size $4.96 \times 4.96$ arcmin$^2$),
at similar galactic latitude to our study ($l \sim 50^{\circ}$), is close to
that expected for a homogeneous Poisson distribution and that the variations of star counts are far lower than those of the galaxy
counts. Considering that there are about three times as many galaxies than
stars at the limiting magnitude ($B_{J}=22.5$) the assumption about the
uniformity of the star distribution should be a good approximation.

\section{FINDING GALAXY CLUSTERS}

Galaxy clusters are the most massive gravitationally bound objects known, and
are the obvious tracers of the LSS. They appear as peaks in the number
density distribution of galaxies. Since we are interested in the position of quasars with respect to cores of galaxy clusters and want to resolve the subclusters if present, it is
crucial to preserve the topological information when sampling the density
distribution. The Voronoi tessellation has been chosen, since it is a
non-parametric method. Voronoi tessellation, also known in computational geometry as
Dirichlet tessellation, provides a partition of a point pattern according to
its spatial structure. Features of this kind of decomposition can also be
used for analysis of the underlying point process and provide a
non-parametric approach to complex point processes, such as galaxy
clustering.

Given a set $\bmath{S}$ of $n$ distinct points in $\cal R$$^{d}$, the
Voronoi diagram is the partition of $\cal R$$^{d}$ into $n$ polyhedral
regions $a(p)$ for $(p \in \bmath{S})$. Each region of $a(p)$, called the Voronoi
cell of $p$, is defined as the set of points in  $\cal R$$^{d}$ which are
closer to $p$ than any other points in $\bmath{S}$:
\\
\\
$a(p) = \{x \in  \cal R$$^{d} ~\mid~ \bar{r}(x,p) \le \bar{r}(x,q) ~~\forall~~ q \in \bmath{S} - p\}$
\\
\\
where $\bar{r}$ is the Euclidean distance function (Okabe et al. 2000 and
references there). One can use different distance functions to define various
variations of Voronoi diagrams, but they will not be discussed here. The set
of all Voronoi cells and their faces forms a cell complex. The vertices of
this complex are called the Voronoi vertices, and the extreme rays
(i.e. unbounded edges) are the Voronoi edges. 

A useful application of the Voronoi model is the compilation of surface density
maps from point data. If one generates the Voronoi diagram of the point set
$\bmath{S}$ and measures the area $A_{i}$ of each polygon $a_{i}$, one can
consider $A_{i}^{-1}$ as an indicator of the local intensity of the point
pattern at $a_{i}$ (Duyckaerts, Godefroy \& Hauw 1994). Using the astrophysical
nomenclature, $A_{i}$ is the local number density at the position of the object
$i$ from the sample.

The Voronoi tessellation diagrams  permit the generation of  contour plots of the density distribution without gridding the data. Gridding of data introduces,
in many cases, undesired smoothing and parameterisation, which distort the
morphology of the measured distribution.

The Voronoi tessellation has been used successfully for finding galaxy
clusters (Ramella et al. 2001, Kim et al. 2000) and its performance compared
with that of the matched filter method by Ramella et al. (2001). In contrast
to Ramella et al. (2001), we detect galaxy clusters by thresholding of the
density contours in colour slices and not in magnitude slices.
The use of colour slices has been proposed and tested by Gladders \& Yee
(2000). Selecting galaxies in colour slices in the colour-magnitude diagram
increases the density contrast of any grouping of early-type galaxies,
which are normally found in clusters or groups. A colour slice in the
colour-magnitude plane provides separation in redshift space via the red
sequence of early-type galaxies in clusters (Gladders \& Yee 2000, and
references inside), thus reducing considerably the projection effects.

The redshift range $0 < z < 0.4$ corresponds to a $B_J - R$ colour space in
the range $1.4 < B_J - R < 2.5$ using the galaxy colours derived by Fukugita,
Shimasaku \& Ichikawa (1995). The limit in the redshift space for
this pair of filters is $z \sim 0.4$, since the colours for higher redshift ellipticals start
to drop for $z > 0.6$, reaching $B_J - R \sim 2.5$ again at $z \sim
0.8$. Such high redshift clusters are in any case beyond the magnitude limit of our data. Using the colours of the known clusters and the model predictions (Kodama et al. 1998) the appropriate gradient of the slices in the
colour-magnitude plane has been deduced to be $-0.06$. The width of the slices
is mainly dictated by the uncertainities in the photometric calibration and
empirically selected to be $\Delta (B_J - R) = 0.2$ mag. The colour slices are
overlapping in steps of 0.1 mag.

The analysis is performed initially in every colour slice separately. The points belonging to candidates for galaxy clusters are selected as those exceeding an intensity threshold $\frac{2}{n} \sum_{i=1}^{n}
a_i^{-1}$, where $a_i$ is the area of the $i^{th}$ polygon (corresponding
to the $i^{th}$ object) and $n$ is the total number of points (polygons). This threshold corresponds to $(N_{field} +
N_{cluster})/N_{field} \sim 2$ where $N_{field}$ is the number density of
field galaxies integrated in the line of sight to $z=0.3$ and
$N_{cluster}$ is the number density of the core of a poor cluster. The
literature values for the densities have been adopted from Cox
(1999). Figure 1 illustrates the distribution of the densities in a
typical colour slice, and the selected threshold, which usually picks $\sim
5\%$ of the cells in the slice.

\begin{figure}
\psfig{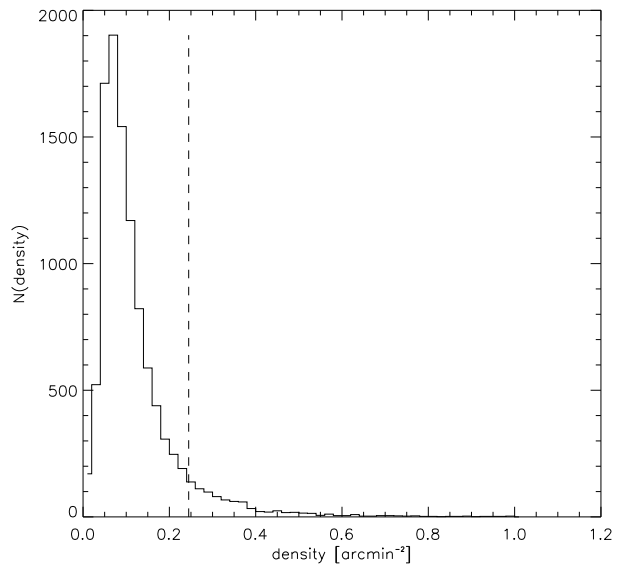}
\caption{The histogram illustrates the distribution of the Voronoi densities in
a typical colour slice (in this example $1.9 < (B_J - R) \le 2.1$). The
dashed vertical line indicates the threshold density.}
\label{fig1}
\end{figure}

\setcounter{figure}{1}
\begin{figure*}
\psfig{file=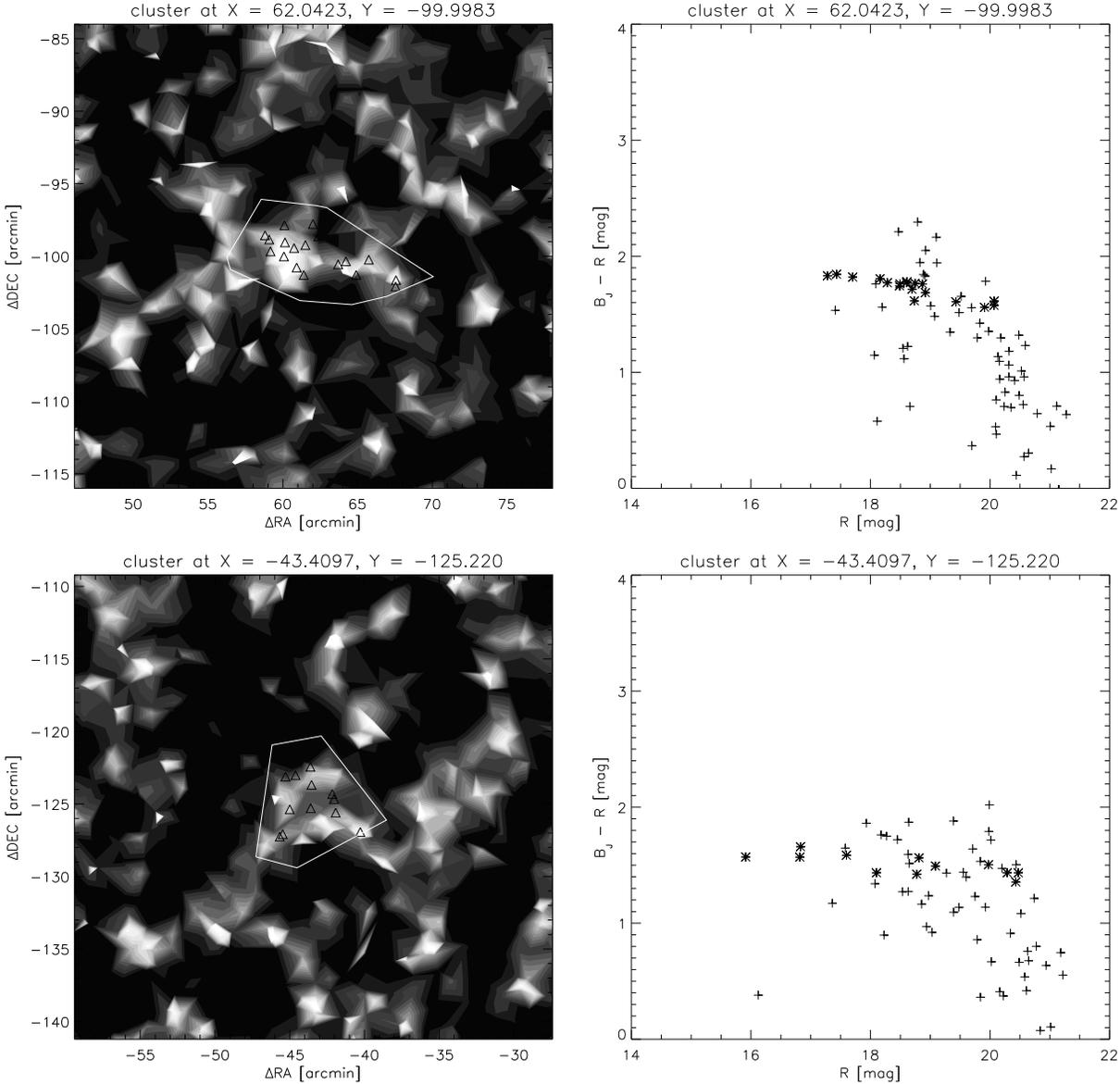,width=16.0cm}
\caption{Two examples of detected clusters. For each cluster, the panel on
the left is a contour plot of the density distribution, with the polygon
enclosing the core of the cluster with early-type galaxies (triangles). The
panel on the right is a colour-magnitude diagram of all objects contained in
the polygon, with those objects indicated by asterisks defining the cluster
detection.}
\label{fig2}
\end{figure*}

The colour-magnitude relation applies to the elliptical galaxies which in turn are known to form the cores of the clusters. The cluster cores can be
expected to be mathematically closed (no holes or single-object dips in the
boundary) structures. Strongly non-virialised clusters, such as Virgo, consisting of multiple subclusters, may be considered as associations of multiple separated cores. The algorithm employed to close the boundaries of the cores is based on the principle that detected structures are closed if all cells with a simple majority of its natural neighbours (adjacent cells) exceeding the density threshold are being included in the selection.  

The Voronoi tessellation has the power to detect groupings on the
richness scale from loose groups to rich clusters. To find only groupings
which qualify as clusters, a lower limit on the number of interconnected
objects in a cluster must be introduced. The lower limit for a poor cluster
(Abell class 0) is by definition 30 galaxies. To detect
such poor clusters with an ellipticals content as low as 25 per cent a minimum of 7
interconnected objects is required.  The results from all the colour slices
have been combined according to the principle that two clusters from
different slices sharing two or more members constitute the same cluster.

\section{CLUSTER SAMPLE}

\setcounter{table}{1}
\begin{table*}
\begin{minipage}{120mm}
\caption{Cluster sample.}
\label{tab2}
\begin{tabular}{ccrrcccrrc}
  RA    &  Dec   &  X   &  Y   &area& $N_G$ & $R_P$ &  $z$   & $N_R$ & ID   \\
hh mm   & dd mm  & arcmin & arcmin & arcmin$^2$ &         & arcmin  &        &       & Zwicky \\
\hline
 10 31.3   &  +05 16  & $-$168  & 32  & 21 & 31   & 1.9  & 0.22  & 34   &   ---  \\ 
 10 31.5   &  +03 46  & $-$165  & $-$58  & 40 & 46   & 3.3  & 0.07  & 97   &   ---  \\ 
 10 32.1   &  +04 26  & $-$156  & $-$18  & 37 & 44   & 2.8  & 0.22  & 22   &   ---  \\ 
 10 32.4   &  +04 59  & $-$151  & 16  & 43 & 75   & 2.8  & 0.20  & 21   & 3280 \\ 
 10 32.5   &  +02 42  & $-$149  & $-$122  & 25 & 29   & 2.2  & 0.12  & 11   &   ---  \\ 
 10 32.8   &  +02 36  & $-$146  & $-$128  & 26 & 40   & 2.4  & 0.12  & 60   &   ---  \\ 
 10 33.0   &  +02 12  & $-$143  & $-$152  & 56 & 99   & 3.1  & 0.15  & 44   &   ---  \\ 
 10 33.0   &  +05 59  & $-$142  & 75  & 61 & 67   & 3.8  & 0.17  & 33   &   ---  \\ 
 10 33.3   &  +05 12  & $-$139  & 28  & 30 & 45   & 2.9  & 0.17  & 19   &   ---  \\ 
 10 33.4   &  +05 01  & $-$137  & 17  & 27 & 43   & 2.7  & 0.20  & 34   &   ---  \\ 
 10 33.9   &  +04 23  & $-$129  & $-$21  & 35 & 51   & 3.4  & 0.12  & 80   &   ---  \\ 
 10 34.1   &  +05 39  & $-$126  & 55  & 13 & 26   & 2.1  & 0.17  & 35   & 3290 \\ 
 10 34.7   &  +04 26  & $-$117  & $-$18  & 39 & 77   & 3.5  & 0.17  & 47   & 3287 \\ 
 10 34.7   &  +06 11  & $-$116  & 87  & 23 & 57   & 1.7  & 0.22  & 25   & 3304 \\ 
 10 35.0   &  +04 39  & $-$113  & $-$4  & 43 & 50   & 3.1  & 0.07  & 191   &   ---  \\ 
 10 35.1   &  +03 34  & $-$111  & $-$70  & 28 & 45   & 2.4  & 0.07  & 110   & 3303 \\ 
 10 35.3   &  +04 03  & $-$108  & $-$41  & 37 & 44   & 2.9  & 0.15  & 24   &   ---  \\ 
 10 35.3   &  +03 55  & $-$108  & $-$49  & 30 & 23   & 2.8  & 0.17  & 40   &   ---  \\ 
 10 35.4   &  +05 59  & $-$106  & 75  & 32 & 39   & 2.4  & 0.20  & 26   &   ---  \\ 
 10 35.5   &  +05 18  & $-$105  & 34  & 87 & 87   & 4.7  & 0.32  & 24   &   ---  \\ 
 10 36.2   &  +02 50  & $-$95  & $-$114  & 30 & 49   & 2.4  & 0.15  & 24   &   ---  \\ 
 10 36.3   &  +03 11  & $-$93  & $-$93  & 38 & 77   & 2.7  & 0.22  & 10   &   ---  \\ 
 10 36.3   &  +07 21  & $-$93  & 157  & 39 & 52   & 2.6  & 0.15  & 28   &   ---  \\ 
 10 36.6   &  +06 22  & $-$88  & 98  & 25 & 32   & 2.1  & 0.20  & 27   &   ---  \\ 
 10 36.7   &  +07 13  & $-$87  & 149  & 26 & 47   & 2.6  & 0.15  & 40   & 3322 \\ 
 10 37.5   &  +05 46  & $-$75  & 62  & 11 & 10   & 4.7  & 0.32  & 11   &   ---  \\ 
 10 37.6   &  +04 07  & $-$73  & $-$37  & 31 & 42   & 2.8  & 0.22  & 48   &   ---  \\ 
 10 38.3   &  +04 53  & $-$63  & 9  & 44 & 45   & 2.5  & 0.07  & 235   &   ---  \\ 
 10 38.3   &  +04 53  & $-$63  & 9  & 24 & 30   & 2.5  & 0.27  & 235   &   ---  \\ 
 10 38.5   &  +04 59  & $-$60  & 15  & 24 & 36   & 2.7  & 0.07  & 245   &   ---  \\ 
 10 38.6   &  +05 09  & $-$58  & 25  & 30 & 32   & 2.1  & 0.22  & 52   &   ---  \\ 
 10 38.9   &  +04 41  & $-$54  & $-$3  & 45 & 79   & 7.2  & 0.12  & 34   &   ---  \\ 
 10 38.9   &  +04 41  & $-$54  & $-$3  & 25 & 39   & 3.2  & 0.27  & 24   &   ---  \\ 
 10 39.1   &  +04 12  & $-$52  & $-$32  & 87 & 10   & 4.5  & 0.17  & 21   &   ---  \\ 
 10 39.1   &  +02 19  & $-$51  & $-$145  & 39 & 55   & 2.9  & 0.12  & 48   & 3365 \\ 
 10 39.1   &  +05 13  & $-$51  & 29  & 10 & 18   & 5.7  & 0.07  & 111   & 3367 \\ 
 10 39.6   &  +02 39  & $-$43  & $-$125  & 49 & 68   & 4.3  & 0.15  & 17   &   ---  \\ 
 10 39.7   &  +03 47  & $-$42  & $-$57  & 24 & 28   & 2.2  & 0.17  & 76   &   ---  \\ 
 10 39.9   &  +04 03  & $-$39  & $-$41  & 58 & 62   & 3.5  & 0.32  & 23   &   ---  \\ 
 10 40.1   &  +03 42  & $-$36  & $-$62  & 13 & 14   & 5.9  & 0.32  & 13   &   ---  \\ 
 10 40.5   &  +03 19  & $-$30  & $-$85  & 24 & 47   & 2.4  & 0.17  & 22   &   ---  \\ 
 10 40.6   &  +04 24  & $-$28  & $-$20  & 67 & 83   & 4.3  & 0.32  & 23   &   ---  \\ 
 10 40.7   &  +06 50  & $-$27  & 126  & 56 & 59   & 3.8  & 0.37  & 14   &   ---  \\ 
 10 40.9   &  +04 58  & $-$25  & 14  & 28 & 38   & 2.1  & 0.32  & 6   &   ---  \\ 
 10 41.8   &  +02 35  & $-$10  & $-$129  & 30 & 29   & 2.4  & 0.07  & 89   &   ---  \\ 
 10 42.2   &  +07 24  & $-$4  & 160  & 30 & 36   & 2.6  & 0.27  & 31   &   ---  \\ 
 10 43.2   &  +03 40  & 12  & $-$64  & 29 & 34   & 2.6  & 0.15  & 19   &   ---  \\ 
 10 43.8   &  +04 12  & 19  & $-$32  & 45 & 65   & 3.7  & 0.07  & 63   &   ---  \\ 
 10 44.0   &  +04 35  & 21  & $-$9  & 23 & 26   & 2.5  & 0.17  & 36   &   ---  \\ 
 10 44.0   &  +07 05  & 23  & 141  & 52 & 63   & 3.8  & 0.32  & 21   &   ---  \\ 
 10 44.5   &  +05 25  & 30  & 41  & 81 & 77   & 4.9  & 0.32  & 22   &   ---  \\ 
 10 44.5   &  +05 44  & 30  & 60  & 46 & 53   & 3.3  & 0.32  & 16   &   ---  \\ 
 10 44.5   &  +06 05  & 30  & 81  & 35 & 51   & 2.5  & 0.32  & 27   &   ---  \\ 
 10 45.6   &  +04 18  & 46  & $-$26  & 19 & 42   & 1.8  & 0.20  & 45   & 3448 \\ 
 10 46.0   &  +06 40  & 52  & 116  & 63 & 62   & 5.8  & 0.07  & 14   &   ---  \\ 
 10 46.6   &  +03 04  & 62  & $-$100  & 61 & 78   & 3.2  & 0.20  & 38   & 3460 \\ 
 10 47.9   &  +05 05  & 81  & 22  & 26 & 41   & 2.4  & 0.03  & 367   & 3477 \\ 
 10 48.0   &  +04 40  & 83  & $-$4  & 52 & 78   & 3.7  & 0.07  & 165   &   ---  \\ 
 10 48.4   &  +06 16  & 88  & 92  & 51 & 42   & 3.5  & 0.12  & 14   &   ---  \\ 
\hline
\end{tabular}
\end{minipage}
\end{table*}
\setcounter{table}{1}
\begin{table*}
\begin{minipage}{120mm}
\caption{Cluster sample. Continued}
\label{tab2b}
\begin{tabular}{ccrrcccrrc}
  RA    &  Dec   &   X    &    Y   &   area     &  $ N_G$ & $R_P$ &  $z$   & $N_R$ &   ID   \\
hh mm   & dd mm  & arcmin & arcmin & arcmin$^2$ &         & arcmin &        &       & Zwicky \\
\hline
 10 48.9   &  +07 13  & 96  & 149  & 43 & 67   & 3.0  & 0.22  & 38   & 3489 \\ 
 10 49.0   &  +05 50  & 97  & 66  & 23 & 32   & 2.1  & 0.22  & 21   &   ---  \\ 
 10 48.9   &  +07 23  & 97  & 159  & 33 & 54   & 2.9  & 0.20  & 21   &   ---  \\ 
 10 49.3   &  +05 10  & 101  & 26  & 24 & 40   & 1.9  & 0.22  & 47   &   ---  \\ 
 10 49.4   &  +03 40  & 103  & $-$64  & 22 & 43   & 2.0  & 0.15  & 48   & 3502 \\ 
 10 49.5   &  +05 54  & 105  & 70  & 49 & 66   & 3.9  & 0.22  & 26   &   ---  \\ 
 10 50.0   &  +05 25  & 112  & 41  & 29 & 47   & 2.1  & 0.22  & 50   &   ---  \\ 
 10 50.1   &  +03 38  & 114  & $-$66  & 23 & 31   & 2.2  & 0.07  & 183   &   ---  \\ 
 10 50.5   &  +04 44  & 120  & 0  & 49 & 62   & 3.7  & 0.17  & 21   &   ---  \\ 
 10 51.3   &  +05 06  & 132  & 22  & 34 & 76   & 3.1  & 0.37  & 21   & 3537 \\ 
 10 51.3   &  +03 41  & 132  & $-$63  & 23 & 29   & 2.3  & 0.07  & 238   &   ---  \\ 
 10 51.7   &  +05 18  & 138  & 34  & 46 & 61   & 2.9  & 0.32  & 22   &   --- \\
 10 51.9   &  +04 58  & 142  & 14  & 55 & 69   & 2.9  & 0.20  & 40   &   ---  \\ 
\hline
\end{tabular}
\end{minipage}
\end{table*}

The application of the above procedure to the $\sim$ 36 deg$^{2}$ area of
investigation produced a sample of 72 galaxy clusters with estimated
redshifts in the range $0.03 \le z \le 0.37$. The resulting density is 2.0
clusters deg$^{-2}$ with $\langle z \rangle = 0.19$. Because of the magnitude
limit, only 13 (optically brightest) clusters with $z > 0.3$ have been
detected. Figure 2 shows some examples of the detected clusters with
different morphologies and redshifts.

The data for all the detected clusters have been compiled in Table 2. The
positions are expressed in RA and Dec (J2000) and as $X$ and $Y$ which are the
coordinates in arc minutes from the field centre (used in all figures). The
area of a cluster core is here defined as the area of the convex hull enclosing all the
Voronoi cells of the early-type galaxies found to be members of the
cluster. It gives a good measure of the core of the cluster, preserving its
morphology. The number of galaxies ($N_G$) in a cluster core is here defined as the background corrected number of
objects found within the convex hull defining the boundary of the core. The
radius ($R_P$) is the minimum radius necessary to enclose the whole convex
hull. The colours of the red sequences have been established
by inspection of the colour-magnitude plots for every cluster. The redshift $z$ has been estimated using the red sequence approach and
the colours of elliptical galaxies compiled by Fukugita et al. (1995). The
uncertainty of the estimate is $\sim 0.05$ due to the uncertainties in the
photometric calibration. The Abell
richness ($N_R$) of the clusters has been computed using the estimated
redshifts to calculate the angular extent of the Abell radius ($1.5h^{-1}$
Mpc). The background counts, used for background corrections, have been taken
as those of a region without any detected clusters. The identifications of
the previously known clusters have been indicated in the final column.

The area of investigation contains 16 published galaxy clusters, 3 of which have known redshifts. We identified 14 of these clusters
including all those with known redshifts. The 2 clusters we could not identify may
be spurious since they were detected only as density peaks in 2D or have few
ellipticals and so are below our threshold in the colour slices. Figure 3 gives the comparison of the published centres of the
clusters and those determined using colour slices. The difference is
strongest for the lowest redshift cluster (X=81, Y=22, $z=0.03$); most of the other
clusters agree very well. The three known redshifts (0.03, 0.07, 0.15)
allowed an improvement of our redshift estimate at $z < 0.2$ by direct
calibration of the colour-redshift sequence.

\begin{figure}
\psfig{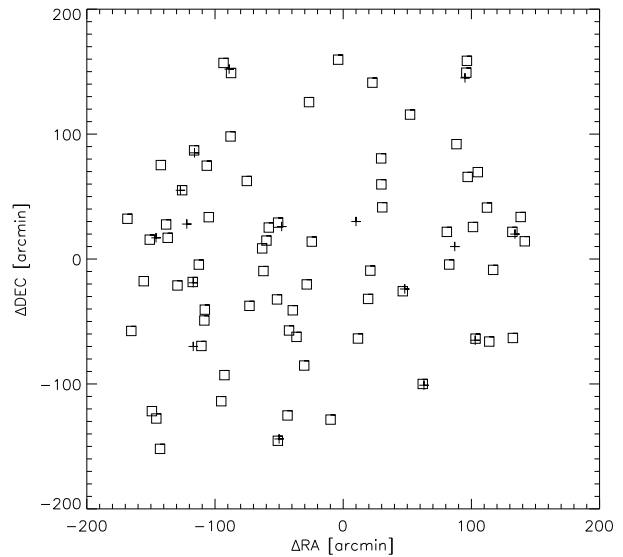}
\caption{Cluster sample. The squares are centred on the galaxy
clusters detected by Voronoi tessellation in ($B_J - R$) colour slices. The
overplotted plus signs represent the known galaxy clusters drawn from the
CEDAG Catalogue of Clusters of Galaxies.}
\label{fig3}
\end{figure}

\section{THE LARGE SCALE ENVIRONMENT OF QUASARS} 

The purpose of our investigation is to establish whether quasars trace the
same structures as the galaxy clusters. The quasar sample covers the redshift
range $0.2 < z < 0.4$. However, the cluster sample appears to be very
incomplete at $z > 0.3$ forcing us to limit the main investigation to one
redshift slice, $0.2 < z < 0.3$. The area of the quasar survey is smaller than that populated by detected clusters.  Only objects in the common
area ($\sim 25$ deg$^2$) of both quasar and cluster samples have been selected for the
environmental study. The final sample comprises 17 galaxy clusters and 7
quasars (Figure 4).

\begin{figure}
\psfig{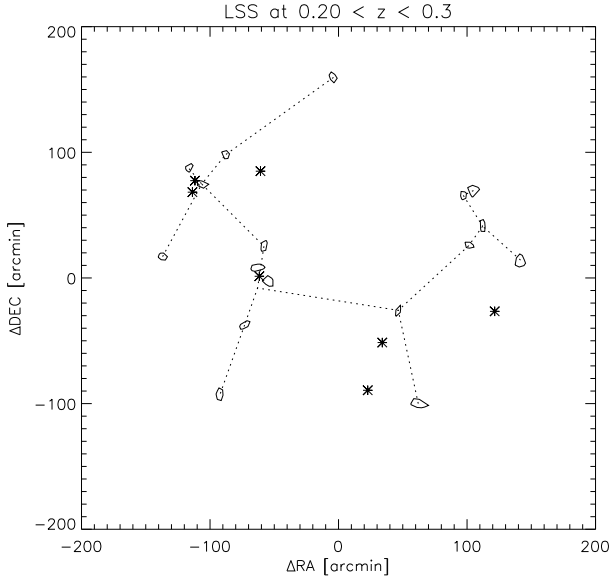}
\caption{The relation between the LSS in galaxy clusters
(polygons) delineated by the MST (dotted line) and quasars (asterisks) in the
redshift slice $0.2 < z < 0.3$. The axes correspond to X and Y in units of
arcminutes from the centre of ESO/SERC field 927.}
\label{fig4}
\end{figure}

The problem of connecting discrete objects into a continuous structure can be
approached using another graph theoretical method, the minimal spanning tree
(MST). This method has been successfully tested for finding LQGs in
quasar samples (Graham 1997). The minimal spanning tree (MST) is a geometric construct, originating in graph theory, which was introduced by Kruskal (1956) and Prim (1957). 

In our case the nodes of the MST are the galaxy clusters. For the purpose of
this study we did not attempt to separate the MST to recover the galaxy
superclusters, but preserved the continuous LSS. When looking at Figure
4, it can be seen that quasars populate the region delineated by the MST as
LSS. In order to test the statistical significance of this result we
resampled the quasar distribution using a Monte Carlo method. As a measure of
how close the quasars follow the LSS, we have used the sum of the distances
between quasars and their closest MST edges $\sum_{i=1}^{n} r_i$. The
result of 100,000 simulations has been plotted as a histogram in Figure 5.
The test rejects the null hypothesis that the quasars are distributed independently of the LSS in clusters at a level of significance of 0.009.

\begin{figure}
\psfig{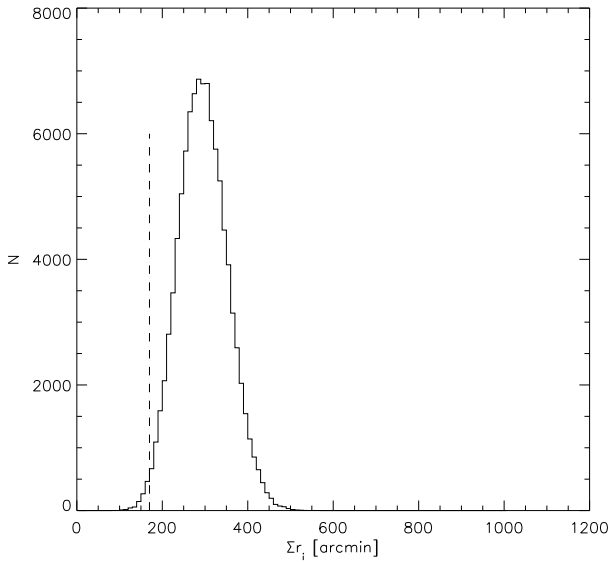}
\caption{The result of the Monte Carlo resampling of the quasar distribution
in relation to the MST vectors. The histogram illustrates the distribution of
$\sum_{i=1}^{n} r_i$, where $r_i$ is the distance of the i$^{th}$ quasar
to its closest MST edge, in 100,000 simulation runs. The vertical dashed
line indicates the value produced by the quasars in the real sample.}
\label{fig5}
\end{figure}

\section{SMALL SCALE ENVIRONMENT OF QUASARS}

From Figure 4, it can be seen that the quasars appear to avoid the central
density peaks of the clusters, and reside instead in two types of
environments. Most of them (5 out of 7) reside on the distant (3 -
5$h^{-1}$ Mpc) peripheries of the galaxy clusters tracing the same LSS. This
is an important result, suggesting that the topology of the density field in
the quasar environments might be a crucial parameter connected with the
triggering of quasar formation and the controlling of their evolution. Some
of the quasars (2 out of 7), however, are found between two very close
clusters. Such an arrangement suggests that quasar formation might also be
associated with cluster mergers.

Figure 6 illustrates the density distribution and the two close clusters around
a quasar at $z = 0.245$. The estimated redshifts of both galaxy clusters are
comparable to that of the quasar, suggesting that all three objects are
physically associated.

\begin{figure}
\psfig{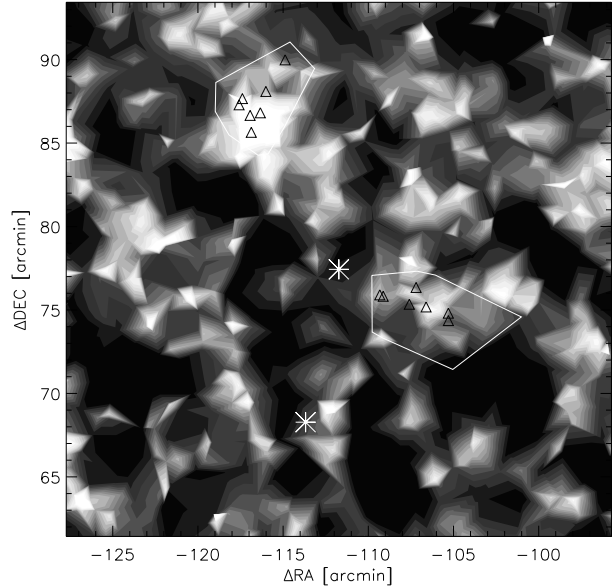}
\caption{The environment of the quasar at $z=0.245$. The figure shows the
density distribution (filled contours) and the cores of two clusters
(polygons) identified at a comparable redshift.}
\label{fig6}
\end{figure}

The second merger candidates are more difficult to verify. Both clusters
found in the vicinity of a quasar at $z = 0.286$ (Figure 7) are obscured by
foreground clusters. They have clear red sequences in the colour-magnitude
diagram but the enclosing polygons and redshift estimates are nevertheless
less reliable than in the first case outlined above. The redshifts of both
clusters are estimated to lie in the range $0.25 < z < 0.30$ which is
compatible with the redshift of the quasar.

\begin{figure}
\psfig{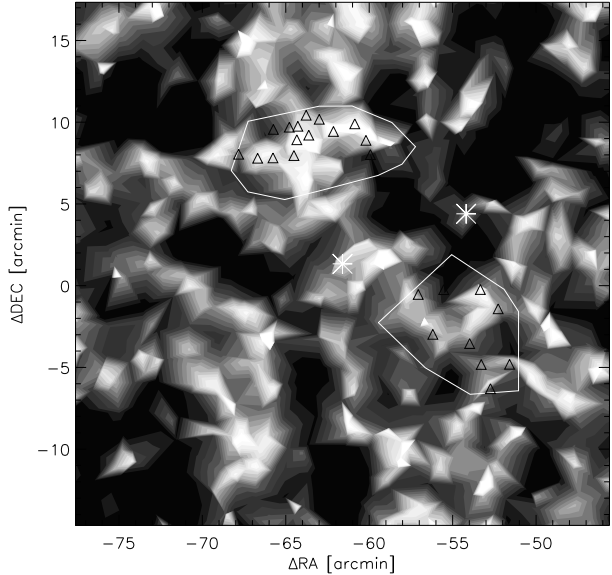}
\caption{The environment of the quasar at $z=0.286$. The figure shows the
density distribution (filled contours) and the cores of two clusters
(polygons) identified at a comparable redshift. The density enhancement
superimposed on the position of the quasar is due to projection and does not
form a cluster core.}
\label{fig7}
\end{figure}

This result suggests that quasars found in rich environments are actually not
in the centres of galaxy clusters but mark the regions of cluster /
subcluster mergers by residing between two (or more) density peaks. With only two quasars, however, this needs to be confirmed with a larger sample (work in progress). 

\section{DISCUSSION AND CONCLUSIONS}

The majority of the studies of the richness of the environments of quasars have relied on counting galaxies within a quite small radius $\sim 1$ Mpc without consideration of the morphology of the distribution. Such an approach provides a fast means of tabulating and comparing the richness within large data sets. However, to ensure a correct interpretation of the results, we must determine what physical structures in the galaxy distribution are associated with quasars. 
We have studied the association of quasars with optically detected clusters of galaxies. By reconstructing the LSS in a relatively narrow redshift band ($0.2 < z < 0.3$), we have been able to show that quasars follow the large-scale structure traced by galaxy clusters. The quasar-cluster LSS investigated in this paper is on the scale of superclusters ($\sim 60h^{-1}$ Mpc), which suggests that quasars could be used to trace the distribution of superclusters. Oort, Arp \& de Ruiter (1981) brought forward for the first time the hypothesis that quasars are located in superclusters using the statistics of close quasar pairs as evidence. It is normally expected that clusters of galaxies trace mass, and that the LSS of galaxies also traces mass. Our results show that quasars trace the mass (galaxies) on supercluster scales for low redshifts $z \la 0.3$. It means that the formation/fuelling process of quasars is coupled, at low redshift, to the large density perturbations, which makes quasars useful indicators of mass at $z \la 0.3$.
We should not, however, simply extend this conclusion to higher redshifts without observational support because the relative importance of different mechanisms for quasar formation will almost certainly vary with redshift.

Secondly, none of the quasars in our radio quiet sample is located in the central area of a galaxy cluster. Apart from the two quasars found between merging clusters, all the others reside on the distant (3 - 5$h^{-1}$ Mpc) peripheries of galaxy clusters. The peripheral position of RQQs in relation to galaxy clusters was first noticed by Yee \& Green (1984) when assessing by eye the distribution of companion galaxies in the vicinity of quasars. Yee (1992) noticed that $z \sim 0.6$ quasars do not reside close to the cD galaxies marking the centres of clusters, but these results do not seem to have been pursued further. Hashimoto \& Oemler (2000) found that galaxy interactions are more likely in galaxy groups and very poor clusters. They found also about the same ratio of interactions in rich clusters as in the field, three times less then in groups and poor clusters. The fact that most quasars in our sample (5 out of 7) avoid strongly the rich environments of the central regions of galaxy clusters is compatible with models connecting the majority of low $z$ quasars with galaxy groups and suggesting that major galaxy-galaxy mergers are the main, but not the only, mechanism of quasar formation at low $z$.

Lastly, we find that quasars found in very rich environment (2 out of 7) may actually be located between two very close galaxy clusters. Due to the relatively small number of quasars in our sample it is not possible to draw conclusions about the percentage of such objects in the overall quasar population. Two clusters separated by a distance $< 2h^{-1}$ Mpc are expected to have entered already the pre-merger stage (Schindler 2001) in which strong gas compression accompanied by heating occurs. The high velocities of galaxies found in merging clusters do not favour major galaxy-galaxy mergers as a quasar formation mechanism, but rather point towards galaxy harassment (Lake, Katz \& Moore 1998). However, cluster /
subcluster mergers produce compression of hot gas between two subclusters
shortly before they collide and shock waves after the collision (Schindler
1999). Considering the complex nature and the high energetics of the gas compressions and displacements already during the pre-merger stage of two galaxy clusters, we should consider also a model in which a quasar is formed in a galaxy-gas interaction. It has been argued that strong star formation can be triggered by the influence of dense ICM at the confluence of a cluster / subcluster merger (Evrard 1991). Our result for $z \sim 0.3$ quasars, combined with the fact that Haines et al. (2001) and Tanaka et al. (2000) found quasars on the peripheries of clusters at $z \sim 1.3$ and 1.1, with some being accompanied by a band of star-forming galaxies and some evidence again for cluster mergers, suggests that such a process could indeed be one of the quasar formation mechanisms. 

To determine the relative contributions of the cluster-merger like and distant periphery like environments to formation of quasars, we are planning to study a large quasar sample, spanning a wide redshift range. The inclusion of radio-loud quasars will allow us to quantify the results as functions of redshift and radio-loudness.
Multiwavelength observations will be required to establish the role of the ICM in triggering the formation of quasars. The X-ray morphology of the merging clusters will give us information about the distribution of the hot gas. Imaging of the host galaxies and their close companions will be used to find traces of recent mergers or interactions.

\section* {ACKNOWLEDGMENTS}

IKS is funded by a bursary of the University of Central Lancashire, UK. LEC
was funded partly by grant 1970/735 from the Fondo Nacional de Desarrollo
Cient\'{\i}fico y Tecnol\'ogico, Chile. We would like to thank Francisco
Castander for constructive discussion about the cluster detection method.

\bsp


\begin{thebibliography}{99}
\bibitem [\protect\citename{Bahcal et al.}1997]{ba}
Bahcall J. N., Kirhakos S., Saxe D. H., Schneider D. P., 1997, ApJ, 479, 642
\bibitem [\protect\citename{Boyce et al.}1998]{bd}
Boyce P. J., Disney M. J., Blades J. C., Boksenberg A., Crane P., Deharveng J. M., Macchetto F. D., Mackay C. D., Sparks W. B., 1998, MNRAS, 298, 121
\bibitem [\protect\citename{Boyle \& Couch}1993]{bc}
Boyle B. J., Couch W. J., 1993, MNRAS, 264, 604
\bibitem [\protect\citename {Clowes }1986] {Clo86}
Clowes R.G., 1986, Mitt. Astron. Ges., 67, 174
\bibitem [\protect\citename {Clowes \& Campusano }1991] {Clo91}
Clowes R.G., Campusano L.E., 1991, MNRAS, 249, 218
\bibitem [\protect\citename {Clowes \& Campusano }1994] {Clo94}
Clowes R.G., Campusano L.E., 1994, MNRAS, 266, 317
\bibitem [\protect\citename {Clowes, Campusano \& Graham }1999] {Clo99}
Clowes R.G., Campusano L.E., Graham M.J., 1999, MNRAS, 309, 48
\bibitem [\protect\citename {Clowes, Cooke \& Beard }1984] {Clo84}
Clowes R.G., Cooke J.A., Beard S.M., 1984, MNRAS 207, 99
\bibitem [\protect\citename {Cox}1999] {Co}
Cox A.N., 1999, editor: Allen's Astrophysical Quantities, 4th edn. AIP Press, Springer 
\bibitem [\protect\citename{Crampton et al.}1987]{cr1}
Crampton D., Cowley A. P., Hartwick F. D. A., 1987, ApJ, 314, 129
\bibitem [\protect\citename{Crampton et al.}1989]{cr2}
Crampton D., Cowley A. P., Hartwick F. D. A., 1989, ApJ, 345, 59
\bibitem[\protect\citename{Duyckaerts, Godefroy \& Hauw}1994]{du}
Duyckaerts C., Godefroy G., Hauw J-J., 1994, J. of Neuroscience Methods, 51, 47
\bibitem [\protect\citename{Ellingson et al.}1991]{el}
Ellingson E., Yee H. K. C., Green R. F., 1991, ApJ, 371, 49
\bibitem [\protect\citename{Evrard}1991]{ev}
Evrard, A.E., 1991, MNRAS, 248, L8
\bibitem [\protect\citename {Fukugita, Shimasaku \& Ichikawa}1995] {Fuk}
Fukugita M., Shimasaku K., Ichikawa T., 1995, PASP, 107, 945
\bibitem [\protect\citename{Gladders \& Yee}2000]{gl}
Gladders M., Yee H.K.C., 2000, AJ, 120, 2148
\bibitem [\protect\citename{Graham}1997]{gr}
Graham M.J., 1997, PhD Thesis, University of Central Lancashire
\bibitem [\protect\citename{Haines et al.}2001]{ha}
Haines C.P., Clowes R.G., Campusano L.E., Adamson A.J., 2001, MNRAS, 323, 688
\bibitem [\protect\citename{Hashimoto \& Oemler}2000]{has}
Hashimoto Y., Oemler A., 2000, ApJ, 530, 652
\bibitem [\protect\citename{Hintzen, Romanishin \& Vlades}1991]{hi}
Hintzen P., Romanishin W., Valdes F., 1991, ApJ, 366, 7
\bibitem [\protect\citename{Hutchings, Crampton \& Johnson}1995]{hu1}
Hutchings J. B., Crampton D., Johnson A., 1995, AJ, 109, 73
\bibitem [\protect\citename{Hutchings at al.}1999]{hu2}
Hutchings J. B., Crampton D., Morris S. L., Durand D., Steinbring E., 1999, ApJ, 117, 1109
\bibitem [\protect\citename{Jones et al.}1991]{jo}
Jones L. R., Fong R., Shanks T., Ellis R. S., Peterson B. A., 1991, MNRAS, 249, 481
\bibitem [\protect\citename {Keable }1987] {Kea87}
Keable C.J., 1987, PhD thesis, University of Edinburgh
\bibitem [\protect\citename{Kim et al. }2000]{ki}
Kim R., et al., 2000, in  Mazure A., Le F\`{e}vre O., Le Brun V., eds., Clustering at High Redshift, ASP Conference Series, Vol. 200, p. 422
\bibitem [\protect\citename {Kodama et al. }1998]{ko}
Kodama T., Arimoto N., Barger A.J., Arag\'{o}n-Salamanca A., 1998, A\&A, 334, 99
\bibitem [\protect\citename{Kruskal}1956]{kr}
Kruskal J. B., 1956, Proc. Am. Math. Soc., 7, 48
\bibitem [\protect\citename{Lake, Katz \& Moore}1998]{la}
Lake G., Katz N., Moore B., 1998, ApJ, 495, 152
\bibitem [\protect\citename{Lim \& Ho}1999]{li}
Lim  J., Ho P. T. P., 1999, ApJ, 510, L7
\bibitem [\protect\citename{McLure et al.}1999]{ml}
McLure R. J., Kukula M. J., Dunlop J. S., Baum S. A., O'Dea C. P., Hughes D. H., 1999, MNRAS, 308, 377
\bibitem [\protect\citename {Newman }1999] {New99}
Newman P.R., 1999, PhD thesis, University of Central Lancashire
\bibitem [\protect\citename {Newman et al. }1998] {New98}
Newman P.R., Clowes R.G., Campusano L.E., Graham M.J., 1998, in Colombi S.,
Mellier Y., Raban B., eds, 14$^{\rm th}$ IAP Colloquium, Wide Field Surveys
in Cosmology. Editions Fronti\`eres, p. 408
\bibitem [\protect\citename{Okabe et al.}2000]{ok}
Okabe A., Boots B., Sugihara K., Chiu S. N., 2000, Spatial Tessellations, 2nd edn. John Wiley \& Sons
\bibitem [\protect\citename{Oort, Arp \& de Ruiter }2000]{oo}
Oort J.H., Arp H., de Ruiter H., 1981, A\&A, 95, 7
\bibitem [\protect\citename{Prim}1957]{pr}
Prim R. C., 1957, Bell Sys. Tech. J., 36, 1389
\bibitem [\protect\citename{Ramella et al. }2001]{ram}
Ramella M., Boschin W., Fadda D., Nonino M., 2001, A\&A, 308, 776
\bibitem [\protect\citename{S\'{a}nchez}1999]{sa}
S\'{a}nchez S.F., Gomz\'{a}les-Serrano J.L., 1999, A\&A, 352, 383
\bibitem [\protect\citename{Saxton, Hall \& Turner}1999]{sa}
Saxton R. D., Hall P. B., Turner M. J. L., 1999, ASP, Vol. 176, p. 389 
\bibitem [\protect\citename{Schindler}1999]{sch}
Schindler S., 1999, astro-ph/9909042
\bibitem [\protect\citename{Schindler}2001]{schb}
Schindler S., 2001, astro-ph/0107008
\bibitem [\protect\citename{Tanaka et al.}2000]{ta}
Tanaka I., et al., 2000, ApJ, 528, 123
\bibitem [\protect\citename{Teplitz, McLean \& Malkan}1999]{te}
Teplitz H. I., McLean I. S., Malkan M. N., 1999, ApJ, 520, 469
\bibitem [\protect\citename{Wold et al.}1999]{wo}
Wold M., Lacy M., Lilje P. B., Serjeant S., 2001, MNRAS, 323, 231
\bibitem [\protect\citename{Yee}1992]{ye}
Yee H.K.C., 1992, in Fabian A.C., ed., Clusters and Superclusters of Galaxies. Kluwer, Dordrecht, p. 293
\bibitem [\protect\citename{Yee \& Green}1984]{yg}
Yee H. K. C., Green R. F., 1984, ApJ, 280, 79
\end{thebibliography}
\end {document}